\documentclass[pra,aps,showpacs,twocolumn]{revtex4}

\usepackage{amsmath}
\usepackage{bm}
\usepackage{graphicx}

\begin{document}

\title{Rayleigh-Taylor instability and mushroom-pattern formation in a
two-component Bose-Einstein condensate}

\author{Kazuki Sasaki$^1$}
\author{Naoya Suzuki$^1$}
\author{Daisuke Akamatsu$^2$}
\author{Hiroki Saito$^1$}

\affiliation{
$^1$Department of Applied Physics and Chemistry,
University of Electro-Communications, Tokyo 182-8585, Japan \\
$^2$National Metrology Institute of Japan (NMIJ),
National Institute of Advanced Industrial Science and Technology (AIST),
Tsukuba 305-8563, Japan}

\date{\today}

\begin{abstract}
The Rayleigh-Taylor instability at the interface in an immiscible
two-component Bose-Einstein condensate is investigated using the 
mean-field and Bogoliubov theories.
Rayleigh-Taylor fingers are found to grow from the interface and
mushroom patterns are formed.
Quantized vortex rings and vortex lines are then generated around the
mushrooms.
The Rayleigh-Taylor instability and mushroom-pattern formation can be
observed in a trapped system.
\end{abstract}


\maketitle

\section{Introduction}

When a layer of a lighter fluid lies under that of a heavier fluid, the
translation symmetry on the interface is spontaneously broken and the
interface is modulated due to the {\it Rayleigh-Taylor} instability
(RTI)~\cite{Rayleigh,Taylor,Lewis,Chandra}.
Waves on the interface then grow into complicated patterns with
mushroom shapes~\cite{Lewis,Daily}.
The RTI plays crucial roles in a variety of nonequilibrium phenomena,
ranging from convection of water in a kettle to supernova
explosions~\cite{supernova}.

In the present paper, we investigate the RTI and ensuing dynamics in a
phase-separated two-component Bose-Einstein condensate (BEC).
Recently, there has been a growing interest in the interface properties
of such BECs.
For instance, the Kelvin-Helmholtz
instability~\cite{Helm,Kelvin,Chandra}, which occurs at the interface
between two fluids with a relative velocity, has been observed in a
$^3{\rm He}$ superfluid system~\cite{Blaa,Volovik}.
The Kelvin-Helmholtz instability is also predicted in a two-component
BEC of atomic gases~\cite{Takeuchi}.
When a magnetic field is applied to a magnetic fluid (a colloidal
suspension of fine magnetic particles), the surface is deformed by the
Rosensweig instability~\cite{Cowley} and grows into a pattern of
crests.
Such a surface phenomenon can be theoretically shown to occur also at
the interface in a two-component BEC with a dipole-dipole
interaction~\cite{Saito}.
Analytical expressions of the interface tension in a two-component
BEC have been derived in Refs.~\cite{Ao,Barankov,Schae}.

The present paper reveals that the RTI emerges at the interface between
two immiscible BECs that are pushed toward each other by, e.g., a
magnetic-field gradient.
We first consider an ideal flat interface, and numerically show that
the interface becomes deformed by the RTI to grow into the well-known
mushroom pattern.
The significant difference between this phenomenon and that in classical
fluids is that the vortices under the caps of the mushrooms are
quantized.
Thus, in three dimensions (3D), quantized vortex rings are generated
around the mushrooms.
Bogoliubov analysis shows that the excitation spectrum of the
interface modes closely resembles that for classical fluids.
We also propose a realistic BEC system in a harmonic trap and show that
the RTI can be observed experimentally.

This paper is organized as follows.
Section~\ref{s:flat} numerically formulates the problem and demonstrates
the RTI and ensuing dynamics for an ideal system.
Section~\ref{s:2DBogo} gives the Bogoliubov spectrum of the interface
modes.
Section~\ref{s:3D} analyzes a trapped system.
Section~\ref{s:conc} provides conclusions to the study.

\section{Rayleigh-Taylor instability at an ideal interface}

\subsection{Mean-field dynamics}
\label{s:flat}

The system considered here is a zero-temperature two-component BEC
described by the Gross-Pitaevskii (GP) equations,
\begin{subequations}
\begin{eqnarray} \label{GP}
i \hbar \frac{\partial \psi_1}{\partial t} & = & \left[
-\frac{\hbar^2}{2m_1} \bm{\nabla}^2 + V_1(\bm{r}) + g_{11} |\psi_1|^2 +
g_{12} |\psi_2|^2 \right] \psi_1, \nonumber \\
\\
i \hbar \frac{\partial \psi_2}{\partial t} & = & \left[
-\frac{\hbar^2}{2m_2} \bm{\nabla}^2 + V_2(\bm{r}) + g_{22} |\psi_2|^2 +
g_{12} |\psi_1|^2 \right] \psi_2, \nonumber \\
\end{eqnarray}
\end{subequations}
where $\psi_j$, $m_j$, and $V_j$ are the macroscopic wave function,
atomic mass, and external potential, respectively, for the $j$th ($j =
1, 2$) component.
The interaction parameters $g_{jj'}$ are given by
\begin{equation}
g_{jj'} = \frac{2 \pi \hbar^2 a_{jj'}}{m_{jj'}},
\end{equation}
where $a_{jj'}$ and $m_{jj'}$ are the $s$-wave scattering length and
reduced mass, respectively, between components $j$ and $j'$.
We assume that the interaction parameters satisfy the phase-separation
condition,
\begin{equation} \label{separate}
g_{11} g_{22} < g_{12}^2.
\end{equation}

For concreteness, we employ the hyperfine states $|F, m_F \rangle = |1,
1 \rangle$ and $|1, -1 \rangle$ of a $^{87}{\rm Rb}$ atom for components
1 and 2, respectively.
According to the scattering lengths reported in Ref.~\cite{Kempen}, we
have $a_{11} = a_{22} = 100.4 a_{\rm B}$ and $a_{12} = 101.3 a_{\rm B}$
with $a_{\rm B}$ being the Bohr radius, and then the condition for the
phase separation in Eq.~(\ref{separate}) is satisfied.
The spin-exchange dynamics $|1, 1 \rangle, |1, -1 \rangle \rightarrow
|1, 0 \rangle, |1, 0 \rangle$ can be suppressed by, e.g., the
microwave-induced quadratic Zeeman effect~\cite{Leslie}, which can lift
the energy of the $|1, 0 \rangle$ state.
When the hyperfine spins are parallel to the magnetic field $\bm{B}$,
the magnetic-field gradient exerts forces $\pm \mu_{\rm B} \bm{\nabla}
|\bm{B}| / 2$ on the two components in opposite directions, where
$\mu_{\rm B}$ is the Bohr magneton.

In order to clearly demonstrate the RTI, we first consider a 3D system
without a trapping potential.
We prepare a quasi-stationary state with a field gradient $B' \equiv dB
/ dz > 0$, which is uniform in the $x$-$y$ direction.
Components 1 and 2 are located in the $z < 0$ and $z > 0$ regions,
respectively, and their flat interface is located at the $z = 0$
plane.
The Stern-Gerlach force pushes the two components toward each other.
We add small random seeds to the initial state, which numerically breaks
the translation symmetry in the $x$-$y$ direction and triggers the RTI.
We assume periodic boundary conditions in the $x$ and $y$ directions.

\begin{figure}[t]
\includegraphics[width=8cm]{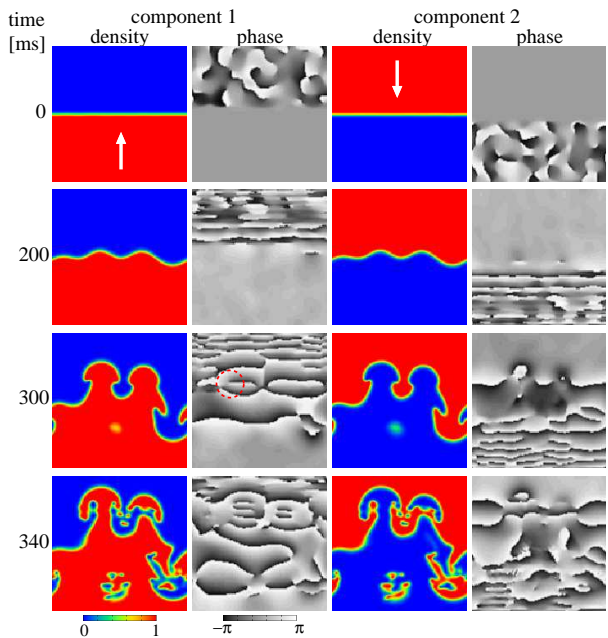}
\caption{
(Color) Density and phase profiles on the $x$-$z$ plane.
The initial state is a quasi-stationary state with a small random seed.
The white arrows show the directions of the Stern-Gerlach force produced
by the field gradient of $B' = 20$ ${\rm mG} / {\rm cm}$.
The red circle indicates the location of a topological defect under a
mushroom cap.
The density is normalized by $4 \times 10^{14}$ ${\rm cm}^{-3}$.
The field of view is $100 \times 100$ $\mu{\rm m}$.
}
\label{f:flat}
\end{figure}
Figure~\ref{f:flat} shows the time evolution of the density and phase
profiles on a plane perpendicular to the initial flat interface.
The two components are pushed toward each other (arrows in 
Fig.~\ref{f:flat}), and the interface starts to modulate due to the RTI
(second row of Fig.~\ref{f:flat}).
Subsequently, the amplitude of the wave on the interface grows to form
the mushroom shapes (third row of Fig.~\ref{f:flat}).
We can see that there are quantized vortices under the caps of the
mushrooms (red circle in Fig.~\ref{f:flat}).
After that, the vortices enter into the mushroom patterns, giving rise
to complicated dynamics (fourth row of Fig.~\ref{f:flat}).

\begin{figure}[t]
\includegraphics[width=7cm]{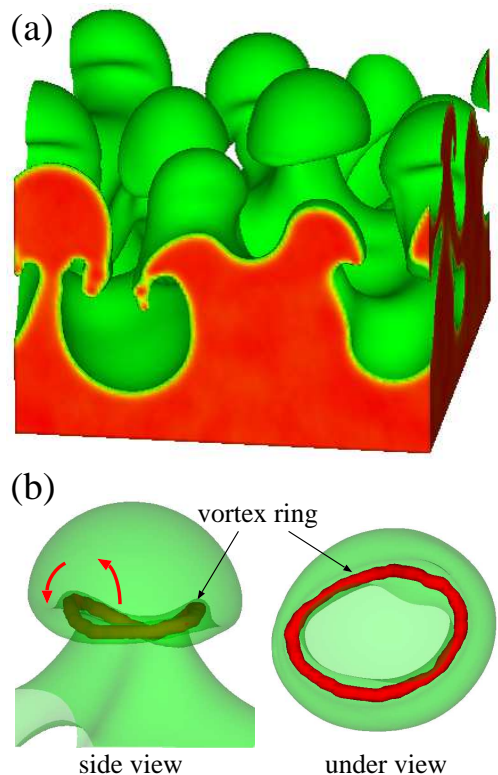}
\caption{
(Color) (a) Isodensity surface of component 1 at $t = 310$ ms.
The condition is the same as that in Fig.~\ref{f:flat}.
(b) One of the mushroom shapes in (a).
The red rings show the location of the topological defects.
The red arrows indicate the directions of the atomic flow.
}
\label{f:3d}
\end{figure}
Figure~\ref{f:3d} (a) shows the isodensity surface of component 1 at $t
= 310$ ms.
We can see that Rayleigh-Taylor fingers and mushroom patterns flourish
at the interface.
One of the mushroom shapes is magnified in Fig.~\ref{f:3d} (b), in which
the topological defect under the cap of the mushroom is indicated by the
red ring.
The vortex rings are generated by the upward flow of the atoms around
the center and downward flow at the periphery of the cap of the mushroom
shape.

\subsection{Bogoliubov analysis}
\label{s:2DBogo}

Before studying the Bogoliubov spectrum, we recall the dispersion
relation for the RTI in classical fluids.
Let us consider a situation in which inviscid incompressible fluids
produce a flat interface perpendicular to the direction of gravity,
where the densities of the lower and upper fluids are $\rho_1$ and
$\rho_2$ with $\rho_1 < \rho_2$.
From the linear analysis, the interface mode has a dispersion relation
of~\cite{Chandra}
\begin{equation} \label{RT}
\omega = \left[ \frac{(\rho_1 - \rho_2) g k + \sigma k^3}{\rho_1 +
					\rho_2} \right]^{1/2},
\end{equation}
where $g$ is the gravitational constant and $\sigma$ is an
interface-tension coefficient.
If gravity is absent, $\omega$ is real for all $k$ and proportional to
$k^{3/2}$.
In the presence of the gravitational force, there is always a range of
$k$ in which $\omega$ is imaginary, and hence the interface is
dynamically unstable.
The range of instability is given by
\begin{equation} \label{kc}
0 < k < \sqrt{\frac{(\rho_2 - \rho_1) g}{\sigma}} \equiv 2\pi
 \lambda_{\rm c}^{-1}, 
\end{equation}
and the most unstable wave number $k_0$ is
\begin{equation} \label{kmu}
k_0 = \sqrt{\frac{(\rho_2 - \rho_1) g}{3\sigma}}.
\end{equation}

We perform the Bogoliubov analysis by decomposing the wave function as
\begin{equation}
\psi_j(\bm{r}, t) = \left[ f_j(z) + \phi_j(\bm{r}) \right] e^{-i
\mu_j t} \qquad (j = 1, 2),
\end{equation}
where $f_j(z)$ is a quasi-stationary state with a flat interface around
the $z = 0$ plane and $\mu_j$ is the chemical potential.
For the case of Fig.~\ref{f:flat}, $f_1(z)$ and $f_2(z)$ are symmetric
with respect to the interface on the $x$-$y$ plane and $\mu_1 = \mu_2
\equiv \mu$ because $g_{11} = g_{22}$.
The small deviation $\phi_j(\bm{r})$ from the ground state is written as
\begin{equation}
\phi_j(\bm{r}) = u_{j,k}(z) e^{i (k x - \omega t)} + v_{j,k}^*(z) e^{-i
 (k x - \omega t)},
\end{equation}
where the wave vector is assumed to be in the $x$ direction without loss
of generality.
The mode functions $u_{j,k}(z)$ and $v_{j,k}(z)$ satisfy the
Bogoliubov-de Gennes equations,
\begin{subequations} \label{bogo}
\begin{eqnarray}
& & \left[ -\frac{\hbar^2}{2m} \left( \frac{\partial^2}{\partial z^2} + k^2
+ V_j \right) - \mu + 2 g_{jj} f_j^2 + g_{jj'} f_{j'}^2 \right] u_{j, k}
\nonumber \\
& & + g_{jj} f_j^2 v_{j, k} + g_{jj'} f_j f_{j'} (u_{j', k} + v_{j', k})
= \hbar \omega u_{j, k}, \\
& & \left[ -\frac{\hbar^2}{2m} \left( \frac{\partial^2}{\partial z^2} + k^2
+ V_j \right) - \mu + 2 g_{jj} f_j^2 + g_{jj'} f_{j'}^2 \right] v_{j, k}
\nonumber \\
& & + g_{jj} f_j^2 u_{j, k} + g_{jj'} f_j f_{j'} (u_{j', k} + v_{j', k})
= -\hbar \omega v_{j, k},
\end{eqnarray}
\end{subequations}
where $(j, j') = (1, 2)$ and $(2, 1)$, $m$ is the mass of $^{87}{\rm
Rb}$, $V_1 = -\mu_{\rm B} B' z / 2$, $V_2 = \mu_{\rm B} B' z / 2$, and
$f_j$ is assumed to be real.

\begin{figure}[t]
\includegraphics[width=9cm]{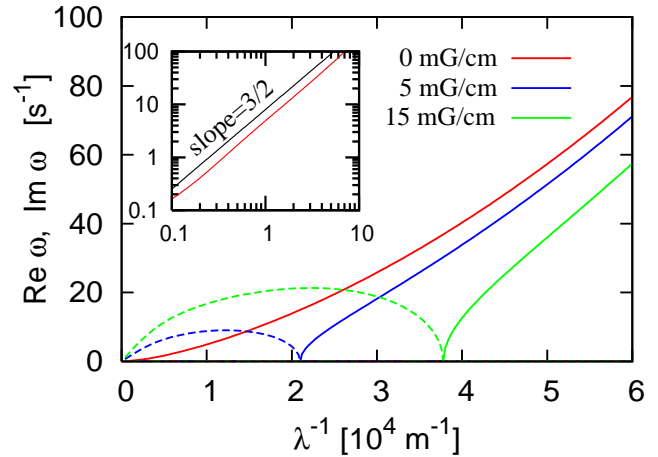}
\caption{
(Color) Real part (solid lines) and imaginary part (dashed lines) of the
Bogoliubov spectrum of an interface mode for a flat interface.
The field gradient perpendicular to the interface is $B' =  0$ (red),
$5$ (blue), and $15$ ${\rm mG} / {\rm cm}$ (green).
The red line in the inset shows a logarithmic plot for $B' =  0$.
For comparison, a black line with a slope of $3/2$ ($\omega \propto
 \lambda^{-3 / 2}$) is shown in the inset.
}
\label{f:bogo}
\end{figure}
Figure~\ref{f:bogo} shows the Bogoliubov spectrum obtained by
numerically diagonalizing Eq.~(\ref{bogo}).
In Fig.~\ref{f:bogo}, we plot only the lowest mode, which corresponds to
the mode localized near the interface.
The second lowest mode has much larger energy.
When the field gradient $B'$ is zero, the excitation energy is real for
all wavelengths $\lambda$ (red line in Fig.~\ref{f:bogo}).
The logarithmic plot in the inset of Fig.~\ref{f:bogo} indicates that
$\omega$ is proportional to $k^{3/2}$, which is in agreement with
Eq.~(\ref{RT}) with $g = 0$.
In the presence of the field gradient $B'$, the Bogoliubov spectrum
becomes imaginary for $\lambda^{-1}$ smaller than a critical value
$\lambda_{\rm c}^{-1}$, which is $\lambda_{\rm c}^{-1} \simeq 2.1 \times
10^4$ ${\rm m}^{-1}$ for $B' = 5$ ${\rm mG / cm}$ and $\lambda_{\rm
c}^{-1} \simeq 3.8 \times 10^4$ ${\rm m}^{-1}$ for $B' = 15$ ${\rm mG /
cm}$.
The long-wavelength modes are always unstable for $B' \neq 0$, as in the
RTI in classical fluids.

The analytic expression of the interface tension in a phase-separated
two-component BEC has been derived in Refs.~\cite{Ao,Barankov,Schae}.
For $a_{12} / a - 1 \ll 1$, where $a \equiv a_{11} = a_{22}$, the
interface tension $\sigma$ has the form,
\begin{equation} \label{sigma}
\sigma = \frac{\hbar^2 n^{3/2}}{m} \sqrt{2\pi(a_{12} - a)},
\end{equation}
where $n$ is the atom density.
Using the characteristic density $4 \times 10^{14}$ ${\rm cm}^{-3}$ for
$n$ and substituting Eq.~(\ref{sigma}) into Eq.~(\ref{kc}), in which
$(\rho_2 - \rho_1) g$ is replaced by $n \mu_{\rm B} B'$, we obtain
$\lambda_{\rm c}^{-1} \simeq 2.1 \times 10^4$ ${\rm
m}^{-1}$ for $B' = 5$ ${\rm mG / cm}$ and $\lambda_{\rm c}^{-1} \simeq
3.6 \times 10^4$ ${\rm m}^{-1}$ for $B' = 15$ ${\rm mG / cm}$.
These values of $\lambda_{\rm c}^{-1}$ are in good agreement with those
in Fig.~\ref{f:bogo}.
Using Eqs.~(\ref{kmu}) and (\ref{sigma}), the most unstable wavelength
is estimated to be $\simeq 40$ $\mu{\rm m}$ for $B' = 20$ ${\rm mG /
cm}$, which is in qualitative agreement with the wavelength of the
interface modulation in Fig.~\ref{f:flat}.

\section{Dynamics in a harmonic trap}
\label{s:3D}

We next consider a system confined in an axisymmetric harmonic potential
$V_{\rm trap} = m [\omega_{xz}^2 (x^2 + z^2) + \omega_y^2 y^2] / 2$.
The radial and axial trap frequencies are $\omega_{xz} = 2\pi \times
100$ Hz and $\omega_z = 2 \pi \times 5$ kHz, and the potential has a
tight pancake shape.
The initial state is the ground state of the GP equation for $B' = 0$,
in which the interface is parallel to the $x$ axis and components 1 and 2
(the hyperfine states $|1, 1 \rangle$ and $|1, -1 \rangle$ of $^{87}{\rm
Rb}$) are localized in the $z < 0$ and $z > 0$ regions, respectively.
A small random noise is added to the initial state to trigger the
dynamical instability.
At $t = 0$, the field gradient $B' = 1.3$ ${\rm G} / {\rm cm}$ is
applied in the $z$ direction.

\begin{figure}[t]
\includegraphics[width=8.5cm]{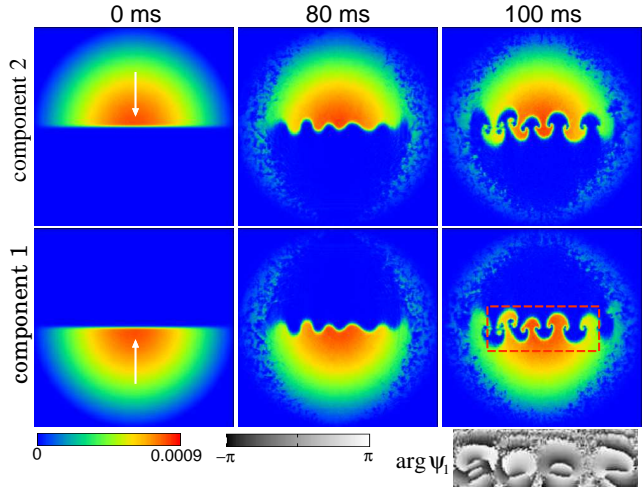}
\caption{
(Color) Column density $\int |\psi_j|^2 dy$ of the two-component BEC in
 a tight pancake-shaped trap with $\omega_{xz} = 2\pi \times 100$ Hz and
$\omega_y = 2\pi \times 5$ kHz.
The number of atoms is $N = 8.1 \times 10^6$ with an equal population in
each component.
The initial state is the ground state for $B' = 0$ plus a small random
seed.
At $t = 0$, the field gradient is changed to $B' = 1.3$ ${\rm G} / {\rm
cm}$.
The white arrows show the directions of the Stern-Gerlach force for the
two components.
The unit of the density is $(m \omega_{xz} / \hbar)^{3/2}$.
The field of view is $64 \times 64$ $\mu{\rm m}$.
}
\label{f:trap}
\end{figure}
Figure~\ref{f:trap} shows the time evolution of the column density $\int
|\psi_j|^2 dy$ of each component, obtained by 3D simulation of the GP
equation.
At $t \simeq 80$ ms, the interface starts to modulate with a wavelength
$\simeq 8$-$10$ $\mu {\rm m}$ due to the RTI.
This wavelength is in qualitative agreement with $\simeq 8.7$ $\mu {\rm
m}$ estimated using Eqs.~(\ref{kmu}) and (\ref{sigma}) with the peak
density $n \simeq 3.2 \times 10^{15}$ ${\rm cm}^{-3}$.
The modulation on the interface then develops into the mushroom patterns
at $t \simeq 100$ ms.
After that, the system evolves in a complicated manner and eventually
components 1 and 2 are interchanged, localizing in the $z > 0$ and $z <
0$ regions, respectively.

In Fig.~\ref{f:trap}, one can see that both components invade around to
the back of each other and the periphery of the condensate is disturbed.
This is because the repulsive interaction between the two components
is weak in the low-density periphery region, and the two components pass
through each other.
For a spherical trap, the pass-through phenomenon is more severe and a
much larger number of atoms ($> 10^9$) is needed to clearly realize the
RTI.
If we use a square-well potential produced by, e.g., a flat-top
beam~\cite{Hao}, the atomic density becomes more uniform on the
interface, which suppresses the pass-through, realizing an ideal RTI as
discussed in Sec.~\ref{s:flat}.
Increasing the inter-component repulsion~\cite{Thal} using the
Feshbach resonance can also suppress the pass-through.

\section{Conclusions}
\label{s:conc}

We have shown that the RTI and mushroom-pattern formation occur in a
two-component phase-separated BEC, as in classical fluids.
The significant difference between the quantum RTI and the classical RTI
is that quantized vortex lines and vortex rings are formed under the
caps of the mushrooms.
The Bogoliubov analysis showed that the excitation spectra of the
interface modes are very similar to those of classical fluids.
We proposed a possible experiment to observe the phenomena in a
realistic trapped system.

Various phenomena related to fluid instabilities may be reproduced in
BECs with renewed interest.
For example, splashing of drops and crown formation~\cite{Yarin} is
considered to be related to instabilities which include the RTI.
Droplet formation by the Plateau-Rayleigh instability~\cite{Rayleigh2}
in BECs is also an interesting future problem.

\begin{acknowledgments}  
We thank S. Tojo for valuable comments.
This work was supported by the Ministry of Education, Culture, Sports,
Science and Technology of Japan (Grants-in-Aid for Scientific Research,
No.\ 17071005 and No.\ 20540388).
\end{acknowledgments}


\begin{thebibliography}{99}

\bibitem{Rayleigh}
Lord Rayleigh, Proc. London Math. Soc. {\bf 14}, 170 (1883).

\bibitem{Taylor}
G. I. Taylor, Proc. Roy. Soc. London Ser. A {\bf 201}, 192 (1950).

\bibitem{Lewis}
D. J. Lewis, Proc. Roy. Soc. London Ser. A {\bf 202}, 81 (1950).

\bibitem{Chandra}
For a textbook, S. Chandrasekhar, {\it Hydrodynamic and Hydromagnetic
Stability} (Clarendon Press, Oxford, 1961), Chapter 10.

\bibitem{Daily}
B. J. Daily, Phys. Fluids {\bf 10}, 297 (1967).

\bibitem{supernova}
A. Burrows, Nature (London) {\bf 403}, 727 (2000).

\bibitem{Helm}
H. von Helmholtz, Phil. Mag. {\bf 36}, 337 (1868).

\bibitem{Kelvin}
Lord Kelvin, Phil. Mag. {\bf 42}, 362 (1871).

\bibitem{Blaa}
R. Blaauwgeers, V. B. Eltsov, G. Eska, A. P. Finne, R. P. Haley,
M. Krusius, J. J. Ruohio, L. Skrbek, and G. E. Volovik,
Phys. Rev. Lett. {\bf 89}, 155301 (2002).

\bibitem{Volovik}
G. E. Volovik, Pis'ma Zh. Eksp. Teor. Fiz. {\bf 75}, 491 (2002)
[JETP Lett. {\bf 75}, 418 (2002)].

\bibitem{Takeuchi}
H. Takeuchi, N. Suzuki, K. Kasamatsu, H. Saito, and M. Tsubota,
arXiv:0909.2144.

\bibitem{Cowley}
M. D. Cowley and R. E. Rosensweig, J. Fluid Mech. {\bf 30}, 671 (1967).

\bibitem{Saito}
H. Saito, Y. Kawaguchi, and M. Ueda, Phys. Rev. Lett. {\bf 102},
230403 (2009).

\bibitem{Ao}
P. Ao and S. T. Chui, Phys. Rev. A {\bf 58}, 4836 (1998).

\bibitem{Barankov}
R. A. Barankov, Phys. Rev. A {\bf 66}, 013612 (2002).

\bibitem{Schae}
B. Van Schaeybroeck, Phys. Rev. A {\bf 78}, 023624 (2008).

\bibitem{Kempen}
E. G. M. van Kempen, S. J. J. M. F. Kokkelmans, D. J. Heinzen, and
B. J. Verhaar, Phys. Rev. Lett. {\bf 88}, 093201 (2002).

\bibitem{Leslie}
S. R. Leslie, J. Guzman, M. Vengalattore, J. D. Sau, M. L. Cohen, and
D. M. Stamper-Kurn, Phys. Rev. A {\bf 79}, 043631 (2009).

\bibitem{Hao}
B. Hao, J. Burch, and J. Leger, Appl. Opt. {\bf 47}, 2931 (2008).

\bibitem{Thal}
G. Thalhammer, G. Barontini, L. De Sarlo, J. Catani, F. Minardi, and
M. Inguscio, Phys. Rev. Lett. {\bf 100}, 210402 (2008).

\bibitem{Yarin}
For review, see, A. L. Yarin, Annu. Rev. Fluid Mech. {\bf 38}, 159
(2006).

\bibitem{Rayleigh2}
Lord Rayleigh, Proc. London Math. Soc. {\bf 10}, 4 (1878);
Proc. R. Soc. London Ser. A {\bf 29}, 71 (1879).

\end{thebibliography}
\end{document}